\documentclass[pre,twocolumn,floatfix,nofootinbib,groupaddress,superscriptaddress]{revtex4-1}
\usepackage{graphicx}
\usepackage{amssymb}
\usepackage{amsmath}
\usepackage{bm}
\usepackage{float}
\usepackage{hyperref}

\begin{document}

\bibliographystyle{apsrev}
%\title{An efficient, learnable, and realizable solution for aquatic bipedalism} 
%\title{Zero-cost gaits and bistability in the dynamics of biomimetic aquatic bipedalism} 
%\title{Benthic bipedalism: from dynamics to robotics} 
%\title{Dynamics and robophysics of benthic bipedalism}
%\title{A theoretical and robotic model for benthic bipedalism}
\title{Models of benthic bipedalism}
%\title{An efficient, learnable, and robotic solution for benthic bipedalism}
%\title{Dynamics and robotics inspired by early benthic bipedalism}
%\title{A robot model suggests efficient and learnable dynamics in early aquatic bipedalism} 
%A simple model for early aquatic bipedalism reveals robust and learnable dynamics in a robotic system
%A robot model suggests robust and learnable dynamics in early aquatic bipedalism
 % A learned gait reveals robust and efficient dynamics of aquatic bipedalism in robots and animals 
% Theoretical and robotic models suggest robust and learnable dynamics in early aquatic bipedalism
% A learned gait and robotic model suggest robust dynamics in early aquatic bipedalism
% Robots mimicking early aquatic bipedalism
% Dynamics and learning of aquatic bipedalism 
% Dynamics of a biomimetic strategy for primal aquatic bipedalism
%(A simple biomimetic strategy for efficient aquatic bipedalism)
%(Evolutionary dynamics and control of aquatic bipedalism)}
\author{F. Giardina}
%\email{}
\affiliation{John A. Paulson School of Engineering and Applied Sciences, Harvard University}
\author{L. Mahadevan}
\affiliation{John A. Paulson School of Engineering and Applied Sciences, Harvard University}
\affiliation{Department of Physics, Harvard University}
\affiliation{Department of Organismic and Evolutionary Biology, Harvard University}

\begin{abstract}
Walking is a common bipedal and quadrupedal gait and is often associated with terrestrial and aquatic organisms. Inspired by recent evidence of the neural underpinnings of primitive aquatic walking in the little skate \textit{Leucoraja erinacea}, we introduce a theoretical model of aquatic walking that reveals robust and efficient gaits with modest requirements for body morphology and control. The model predicts undulatory behavior of the system body with a regular foot placement pattern which is also observed in the animal, and additionally predicts the existence of gait bistability between two states, one with a large energetic cost for locomotion and another associated with almost no energetic cost. We show that these can be discovered using a simple reinforcement learning scheme. To test these theoretical frameworks, we built a bipedal robot and show that its behaviors are similar to those of our minimal model: its gait is also periodic and exhibits bistability, with a low efficiency gait separated from a high efficiency gait by a "jump" transition.  Overall, our study highlights the physical constraints on the evolution of walking and provides a guide for the design of efficient biomimetic robots. 

\end{abstract}

%\date{\today}
%\smallskip

\maketitle

%\large
\section{\label{SIntro}Introduction}
\begin{figure*}
	\centering
	\includegraphics[scale=0.7]{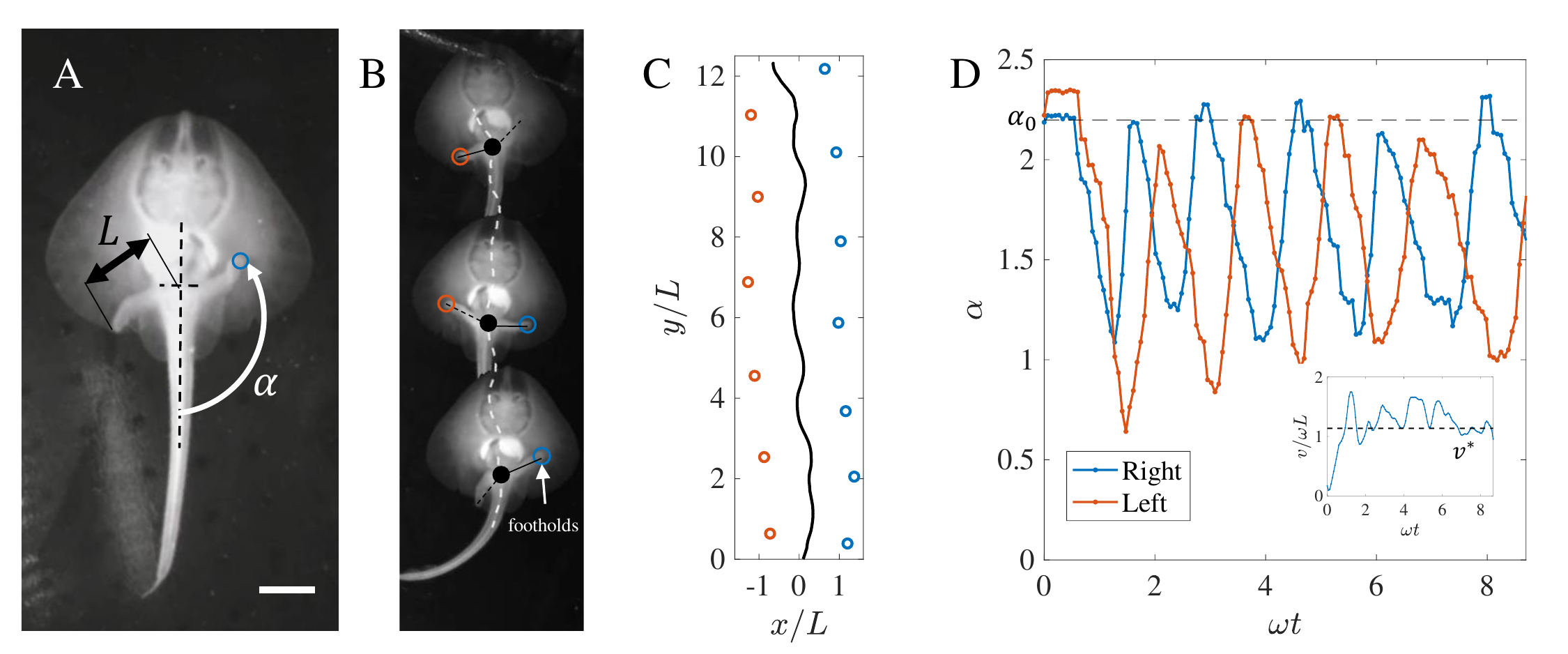}
	\caption{\textbf{Little skate locomotion behavior} (data obtained from \cite{jung2018ancient} with permission from the authors) . \textbf{A}, Ventral view of little skate indicating leg length $L$, footsteps, and leg angle $\alpha$ measured relative to the centerline of the body. Scale bar 1cm. \textbf{B}, Sequence of walking gait indicating trajectory of the pelvic girdle (dashed white line), active legs (solid black lines), passive legs (dashed black lines), and footsteps. \textbf{C}, Trajectory of the pelvic girdle (black line) as a function of dimensionless position with footsteps (circles). \textbf{D}, Left and right leg angles $\alpha$ as a function of dimensionless time and mean foot placement angle $\alpha_0$. The inset shows the dimensionless speed of the pelvic girdle as a function of dimensionless time with $v^*$ (dashed line) the approximate lower speed bound during steady state locomotion.  }
	\label{Fig1}
\end{figure*}

The transition of vertebrates from water to land is thought to have occurred around 400 million years ago and required fundamental changes in morphology and behavior. Used to living in near-neutral buoyancy, aquatic vertebrates had to adapt to the effects of gravity on land which required a change in locomotion strategy. Fins provided a basic form of legs that helped early land-dwelling vertebrates to support their body weight and switch from undulatory (swimming by lateral bending of the body) locomotion to ambulatory (walking) locomotion.  A common view of the transition from undulation to walking is that it occurred gradually \cite{grillner2009measured}, consistent with observations in contemporary tetrapods that most closely represent early legged locomotion. For instance, the salamander uses a combination of undulatory and ambulatory locomotion when walking on land \cite{chevallier2008organisation}, supporting the hypothesis of gradual transition from swimming to walking during vertebrate terrestrialization. While the development of legs can be traced in the fossil record, the origins of neural circuits giving rise to the control required for ambulatory locomotion are unclear. However, recent work \cite{jung2018ancient} suggests that the neural circuits required for limb control can be found in aquatic vertebrates who are distant relatives to the first tetrapods, indicating that the neuromuscular basis for legged locomotion was present in all vertebrates with paired fins. These observations raise the question of  whether a walking gait was actively used in the earliest finned vertebrates or if it only emerged prior to terrestrialization. Given the plethora of extant benthic (living near the bottom of aquatic environments)  fish and species that can walk short stretches on land \cite{king2011behavioral,flammang2016tetrapod,macesic2010comparative,lucifora2002walking}, it is conceivable that their ancestors with similar morphologies might have used these ancient neural circuits to walk. A particularly striking example in this regard is the little skate  \textit{Leucoraja erinacea} (Figs. \ref{Fig1}A,B) that is incapable of undulation due to its rather stiff vertebrae,  and uses a benthic gait consisting of left-right alternating walking \cite{jung2018ancient}. These observations are strongly suggestive that walking in aquatic environments can emerge without a prior undulatory gait and that the a requirement for the evolution of walking is the capability to independently control each fin along with a control strategy that synchronizes the walking motion. While there exists strong evidence that independent leg control was present in early vertebrates with paired fins \cite{jung2018ancient}, what form the locomotion gait adopted and whether a neural control strategy for stable and efficient legged locomotion was feasible remain open questions. %This raises the question: given a benthic near-neutrally buoyant environment and the neuromuscular system present in early vertebrates with leg-like appendages, what gait is likely to be selected?

Here, inspired by published video data on the dynamics of walking in the little skate, we study these questions by devising a minimal mathematical model  to analyze the stability, energy efficiency, and control complexity of early benthic locomotion. We compare the most efficient gaits predicted by the model to the kinematics of the little skate and show that both are characterized by a left-right alternating leg pattern with an undulating center of mass and a regular foot placement pattern. Closed-form expressions for the dynamics of the model show that the most energy efficient gait is associated with no energetic cost of locomotion and merely requires a simple open-loop control strategy. Additionally, the model also predicts the co-existence of a second gait with much lower efficiency. To complement this explicit dynamic model, we use a reinforcement learning strategy and show that a little skate-like gait is the preferred solution in this framework, suggesting that an evolutionary process can converge to this in nature. Finally, to test our results with hardware, we build a simple bipedal robot and show that it also exhibits bistable behavior for certain control parameters, qualitatively consistent with our theory. 

\section{\label{SIntro}Mathematical Model}
Published videos of the little skate \cite{jung2018ancient}  allowed us to extract foot placements, trajectories, and leg angles as shown in Figs. \ref{Fig1}C,D (See also Supplementary Information, SI, and SI Video). In steady locomotion, we observed that as the pelvic girdle position of the skate undulates during locomotion, only one foot is in contact with the ground at any time, and there is little slip between the leg and the ground during contact. This leads to a periodic foot placement pattern over time as shown in Fig. \ref{Fig1}C, accompanied by periodic dynamics of the leg angle $\alpha$ (measured with respect to the centerline of the body), as shown in Fig. \ref{Fig1}D. 
%We observed that the motion occurs in the transverse plane, the foot placement angle $\alpha_0$ stays constant throughout locomotion, the frequency of leg oscillation $\omega$ is approximately constant, and that legs are in opposite phase during steady walking. 

\begin{figure}
	\centering
	\includegraphics[scale=0.8]{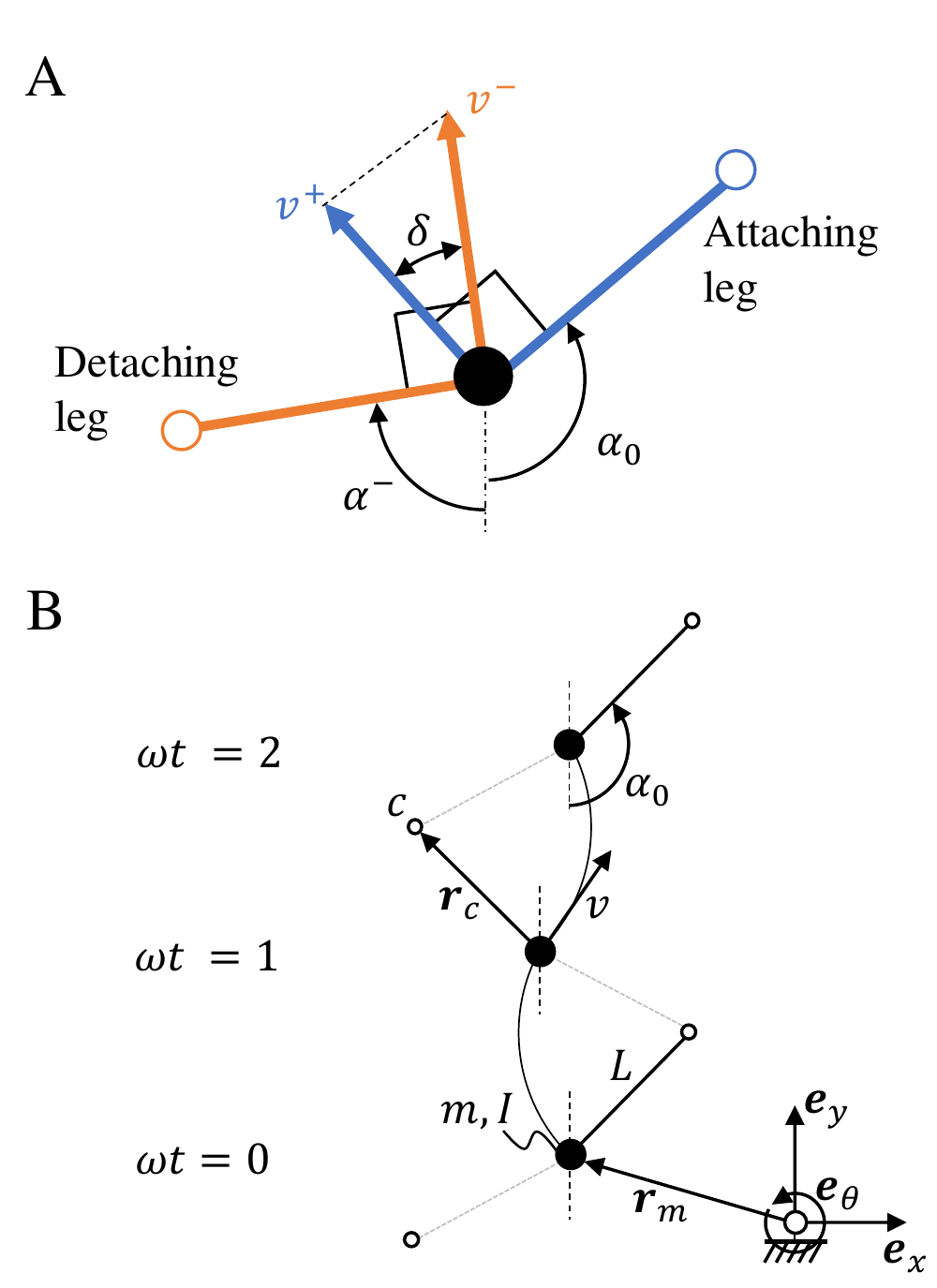}
	\caption{\textbf{Model sketch.} \textbf{A}, The leg transition is modelled with an inelastic impact where the post-impact speed $v^+$ is obtained by a projection of the pre-impact speed $v^-$ to the line perpendicular to the attaching leg direction. \textbf{B}, Model sketch at three subsequent dimensionless times. Point mass $m$ with moment of inertia $I$ is constrained in its motion by a connected massless leg with length $L$ to rotate around current active contact point $c$. The system cannot slip and velocity losses can only occur due to inelastic impact at leg transition. A constant torque $T$ applied to the leg can accelerate the system. The leg placement angle is given with $\pm\alpha_0$ at transition.}
	\label{Fig2}
\end{figure}

Based on these observations,  a minimal model of locomotion suggests modeling the body as a mass $m$ with moment of inertia $I$ which can move and rotate freely in the plane if no legs are placed on the ground, as shown in Fig. \ref{Fig2}; assuming perfect neutral buoyancy removes the effect of gravity. Since the legs are very light relative to the massive body, we ignored their mass, assumed them to be directly attached to the center of mass, and allowed them to switch their ground contact state with a frequency $\omega$. In the absence of slip, upon leg contact with the ground, the velocity of the body $v$ is perpendicular to the leg vector $\bm{r}_c$ which points from body to contact point $c$.  The foot-ground contact was modeled as a perfectly inelastic impact which dissipates any velocity component that violates the constraint at foot placement, consistent with a range of previously used simple models of locomotion \cite{garcia1998simplest,goswami1998study,usherwood2003understanding}. During the impulse-free phases, a torque $T$ accelerates the body on a circular path around the active leg with length $L$ and the current contact point $c$. For simplicity, we assumed the torque to be constant during step $i$, i.e. $T_i(t)=T_i$. 

We described the resulting dynamics of this model  in terms of generalized coordinates $\bm{q}=[x,y,\theta]^T$, with $x,y$ defining the position in the plane and $\theta$ the body angle with respect to the inertial frame of reference. The rotational degree of freedom was assumed to be decoupled from translation during a single step, but coupling occurs at leg transition as the touchdown angle $\alpha_0$ is measured relative to body orientation. In an alternating leg sequence with constant frequency $\omega$ and torque $T$, the body rotates during a step $i$ by $\Delta \theta_i=\pm T/2I\omega^2 + \dot{\theta}_{i-1}/\omega$, where $\dot{\theta}_i= \dot{\theta}_{i-1} \pm T/I\omega$. It is easy to verify that the initial conditions $\dot{\theta}_0= \mp T/2I\omega, \theta=0$ (positive sign for right leg, negative for left) guarantee at every step $\theta_i=0$. The system is subject to opening and closing bilateral contacts with the ground, which is a problem extensively studied in nonsmooth mechanics \cite{brogliato1999nonsmooth}. We modeled a closing contact with an inelastic impact law and assumed an initial condition for $\theta_i=0, \forall i$ as described above. We considered the transition from active ground contact in the left leg to the right leg only, as the final result is leg independent. The center of mass velocity before impact at step $i$ is denoted by $v^-_i$ and after impact with $v_i^+$. The direction of the velocities is given by the geometry of the problem as shown in Fig. \ref{Fig2}A. Pre- and post-impact velocities must be perpendicular to the leg direction (defined by the angles $\alpha^-$ and $\alpha_0$ for detaching and attaching leg, respectively) due to the pivoting motion. At leg transition, the constraints of both legs may not be compatible, which requires a projection of the velocity of the detaching leg to the direction perpendicular to the attaching leg. This projection is the inelastic impact law at step $i$ and is defined as
\begin{equation}\label{eq:impactMap}
v^+_i = \left| \cos \delta_i \right| v^-_i.
\end{equation}
$\delta_i$ is the difference in leg angles at the transition given by $\delta_i=\alpha_0 - \alpha^-_i - \pi$ as shown in Fig. \ref{Fig2}A. Therefore, the mapping of velocity magnitude from pre- to post-impact state is only a function of the leg angles at collision. With the assumption of constant torque $T$ during contact phase, the gained velocity due to torque $T$ over the period $1/\omega$ is $\Delta v=T/\omega L m$. The post-impact velocity map follows 
\begin{equation}\label{eq:velocityMap}
v^+_i = \left| \cos \delta_i \right| \left( v^+_{i-1} + \Delta v \right)
\end{equation}

Equation (\ref{eq:velocityMap}) contains implicitly $\alpha^-_i$ in $\delta_i$ which itself depends on $v^+_{i-1}$ by the relation $\alpha^-_i= -\alpha_0 + (v^+_{i-1}/\omega L + \gamma/2)$, with the nondimensional torque $\gamma=T/\omega^2L^2m$. This is the evolution of $\alpha$ over one step given the initial velocity and prescribed torque.

The kinematic data in Fig. \ref{Fig1}D and SI suggests that, after a short transient phase, the leg torque $T$, frequency $\omega$, and leg placement angle $\alpha_0$ are constant over consecutive steps. This observation suggests the following open loop \textit{bipedal control strategy}: alternate in leg activation and keep $T$, $\omega$ and $\alpha_0$ constant over all steps. This allowed us to frame the energetic cost, speed, and stability of the locomotion gait mathematically. Since all control parameters are constant, the speeds at which loss due to collision and gain due to torque $T$ balance and result in no variation in speed over subsequent steps, i.e. the fixed points of the system, are given by the implicit equation

\begin{equation}\label{eq:FixedPoint}
\bar{v}^* \left( 1 - \left|\cos{\delta}\right| \right) = \gamma \left|\cos{\delta}\right|,
\end{equation}

with $\bar{v}^*=v^*/\omega L$. The evolution of the fixed point corresponds to a Poincar\'e map of the dynamics of the studied body. For $v_0=v^*$ the resulting trajectory of the dynamical system is a limit cycle with period $2/\omega$, where the prefactor is due to the bilateral symmetry. Continuous undulation of the body and a regular footstep pattern are the prominent features of the gait, which are also observed in recorded animal data shown in \ref{Fig1}B. As mentioned above, the derivation of (\ref{eq:FixedPoint}) does not include dynamics of the body rotation because we can decouple rotation from translation with the initial rotational velocity $\pm T/2\omega I$ and initial body orientation $\theta_0=0$.

Stability of the discrete map (\ref{eq:velocityMap}) was quantified by applying a linear stability analysis, i.e. searching for $v^*$ such that

\begin{equation}\label{eq:stability}
\left| \frac{\mathrm{d} v_i^+}{\mathrm{d} v_{i-1}^+} \right|_{v_{i-1}^+=v^*} <1.
\end{equation}

%\section{\label{SIntro}Analysis of Model}

\begin{figure}
	\centering
	\includegraphics[scale=0.7]{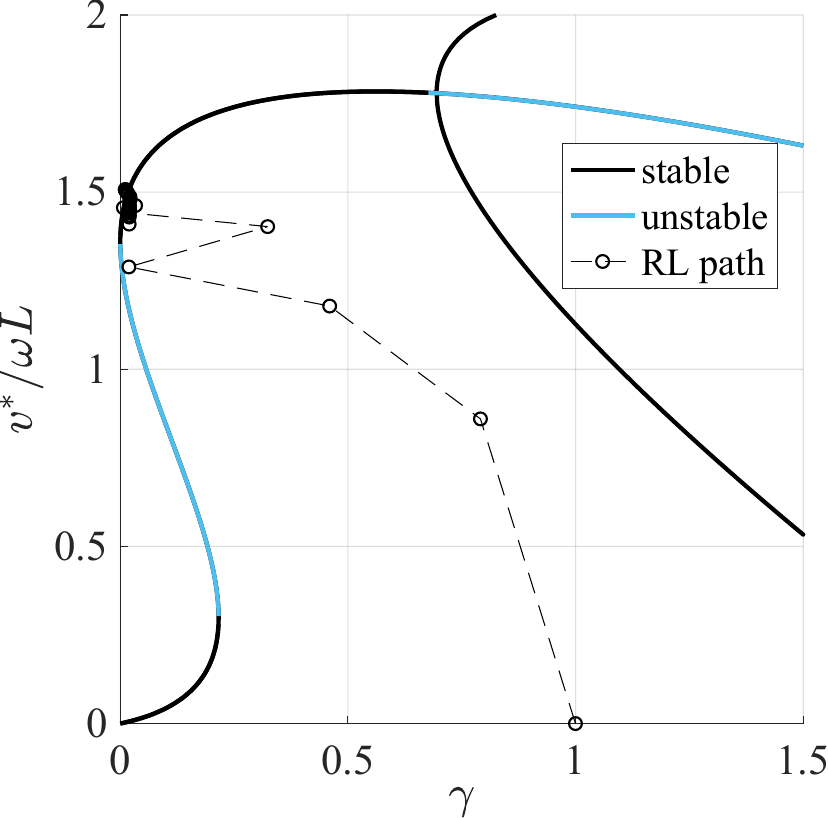}
	\caption{\textbf{Dynamics of bipedal locomotion strategy.} Bifurcation diagram with nondimensional torque $\gamma=T/\omega^2L^2m$ for a fixed foot placement angle $\alpha_0=2.25$. Black lines are stable fixed-point velocities $\bar{v}^*=v^*/\omega L$ and blue lines unstable ones. The bifurcation around  $\gamma\approx0.75$ is a period-doubling bifurcation and the solution jumps between the lower and upper black branch. Black circles show consecutive steps of the optimized reinforcement learning policy.}
	\label{Fig3}
\end{figure}

To reveal the possible behaviors and gaits of the model we searched for solutions of the discrete map (\ref{eq:FixedPoint}). Finding a solution (i.e. a fixed point $\bar{v}^*$) in the model equates to finding a periodic gait.  For a constant foot placement angle $\alpha_0$ the fixed point velocity $\bar{v}^*$ is solely defined by the nondimensional torque $\gamma$. Fig. \ref{Fig3} shows the solutions of (\ref{eq:FixedPoint}) in a bifurcation diagram for a fixed foot placement angle $\alpha_0$.  Stability analysis of the solutions revealed there are three stable regions of interest in the diagram. For $\gamma \le 0.2$ we found two solution branches that coexist, one at low speed and the other one at higher speed connected by an unstable region. A system on the edge of the lower branch will experience a sudden jump in its attracting fixed point as the nondimensional torque $\gamma$ is increased, consistent with the existence of a saddle-node bifurcation \cite{strogatz2018nonlinear}. For  $\gamma \in [0.2, 0.75]$ only one stable fixed point exists which is the continuation of the upper branch. Lastly, at around $\gamma\approx0.75$ a period doubling bifurcation occurs, and the system jumps between two solution branches for $\gamma>0.75$. This gait is asymmetric with a sideway component relative to the body orientation.

Gaits with small $\gamma$ are biologically most plausible as they reduce energetic cost. In fact, there exists a point at $\gamma=0$, on the upper saddle-node bifurcation in the bifurcation diagram, for which the post-impact velocity is nonzero, indicating a gait with zero energetic cost. This was confirmed by considering the kinetic energy over time:  energy fluctuations vanished when $\gamma=0$, which required the leg vectors $\bm{r}_c$ of the left and right leg at transition to be parallel, i.e. $\delta=0$. The speed of this point which corresponds to the energy-optimal gait was studied next. For small $\gamma$, the right-hand side of (\ref{eq:FixedPoint}) is negligible, and we have the solutions $\bar{v}^*=0$ and $\left|\cos{\delta}\right|=1$. The first fixed-point speed provides the trivial solution on the lower branch. For the second solution, recall that $\delta=2\alpha_0-\bar{v}^* - \gamma/2 - \pi$ which implies $\bar{v}^* = 2\alpha_0 - n\pi, n\in \mathbb{Z}$. For $n=1$ the solution corresponds to the energy-optimal gait. The physical interpretation of this result is that for the energy-optimal gait, the angle $\delta$ has to be zero, or put differently, the legs have to be parallel at the transition which removes any dissipation due to impact. Interestingly, the parameter $n$ suggests that there are an infinite number of energy-optimal gaits. In these gaits, the transition from one leg to the next occurs after $\left|n \right|/2$ rotations around the same ground contact point, effectively resulting in a pirouette-like rotation before the leg transition. Although the velocity can become rather large, these solutions are less interesting in the context of our investigation of directed bipedal locomotion and were not further analyzed.

While real systems will have to operate at $\gamma>0$ to ensure stability, the upper branch for  $\gamma<0.2$ is not reachable from rest with the bipedal control strategy, and the system will always converge to the lower, slower and less efficient branch. To switch to the upper branch, one has to literally “leap” onto it using an impulse-like initial actuation strategy. Indeed, such motions are observed in the little skate as it takes off from rest by punting forward with a powerful stroke using both legs at the same time, followed by immediate switching to the alternating gait (See Figs. \ref{Fig1}D, SI). 

\begin{figure}
	\centering
	\includegraphics[scale=0.47]{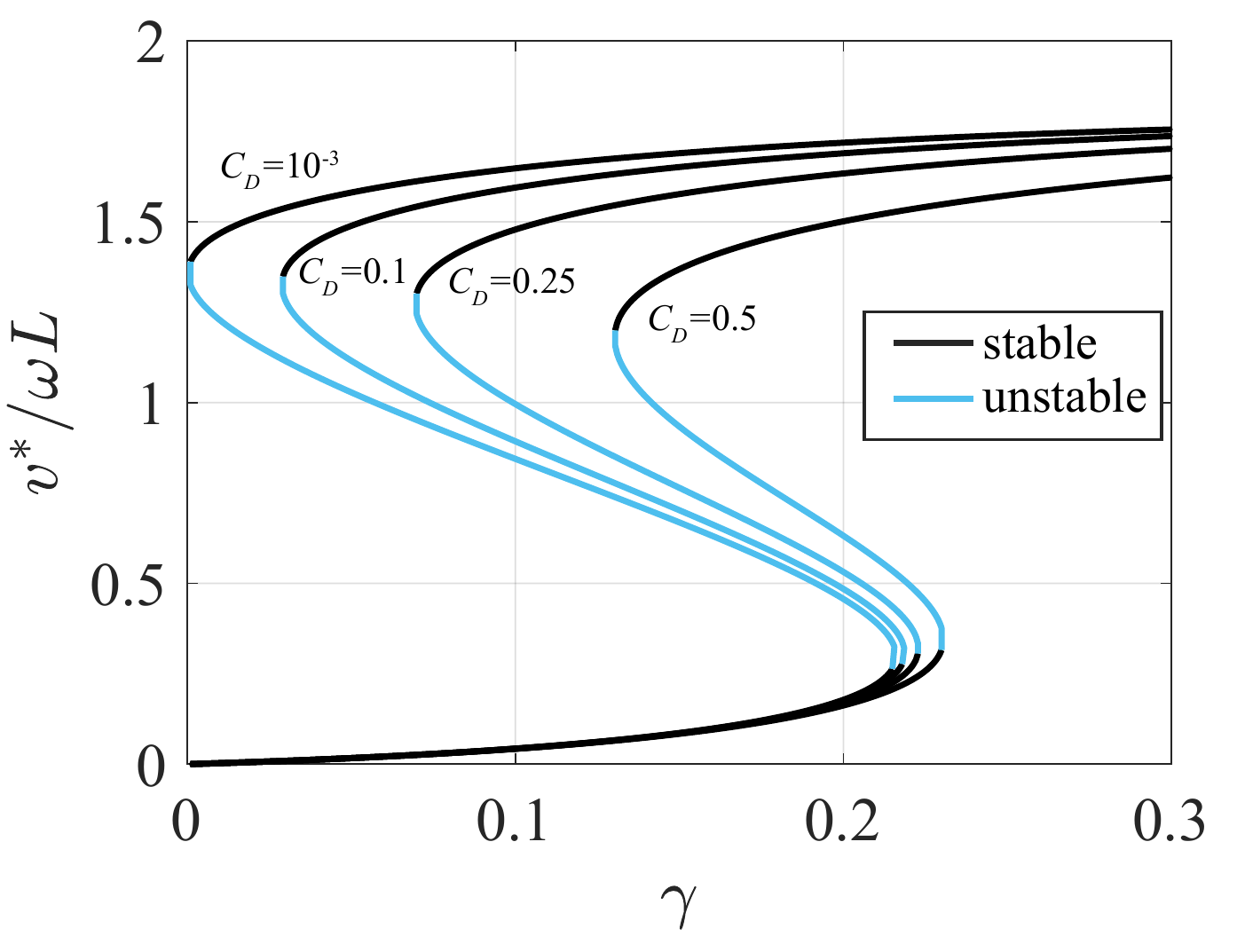}
	\caption{\textbf{Bifurcation diagram with fluid drag for various drag coefficients.} Bifurcation diagram as a function of nondimensional torque $\gamma=T/\omega^2 L^2 m$. Black lines are stable fixed-point velocities $v^*/\omega L$ and blue lines unstable ones. Equations for the fixed points are obtained by using (\ref{eq:dragSol}) in (\ref{eq:velocityMap}). As $C_d \rightarrow 0$  the bifurcation diagram approaches the result shown in Fig. \ref{Fig3}. }
	\label{EFig2}
\end{figure}

The effect of fluid drag on the fixed points  $\bar{v}^*$ was studied next. The Reynolds number in the little skate is of the order $10^3$ and drag can be expressed using the drag equation $F_d=1/2\rho v^2 C_dA$, with $\rho$ the fluid density, $v$ the body velocity, $C_d$ the drag coefficient, and $A$ the reference area. The part in (\ref{eq:velocityMap}) which gets affected by this dissipative effect is $\Delta v$, i.e. the velocity increase over the contact phase of one leg. The velocity in the case with drag obeys the differential equation
\begin{equation}\label{eq:wdrag}
\frac{\mathrm{d} v}{\mathrm{d} t} = \frac{T}{mL} -\frac{1}{2}\rho v^2 C_dA, 
\end{equation}
which has the solution
\begin{equation}\label{eq:dragSol}
v(t) = \sqrt{\frac{2T}{L\rho C_d A}} \tanh \left(\sqrt{\frac{\rho C_d A T}{2 m^2 L}} (c_1 m L + t) \right)
\end{equation}
with $c_1$ a constant to be defined by initial conditions. This provides an analogous closed form of (\ref{eq:velocityMap}) and enables the search for fixed points in the system. The solution depends on the drag coefficient $C_d$ and we expected to retrieve the original drag-less solution for $C_d \rightarrow 0$. Fig. \ref{EFig2} shows the bifurcation diagram for the case with fluid drag for various drag coefficients and reference area estimated from specimen dimensions. For small drag coefficients, the bifurcation diagram converges to the case with no drag. As the drag coefficient increases, the upper stable branch moves towards larger $\gamma$ values and the fixed-point velocities decrease. We do not know the actual drag coefficient for the little skate, but estimations of a benthic ray (Raja clavata) suggest a drag coefficient of 0.1 (see \cite{webb1989station}). At the considered Reynolds numbers the drag coefficient of a sphere in a fluid flow is $C_d \approx 0.5$  and even for this conservative case we found bistable parameter values and similar converged fixed-point speeds on the lower and upper branches compared to the no-drag case, although the nondimensional torque $\gamma$ required for a similar speed increases significantly. Around $C_d \approx 1.08$ the bistable region disappears completely and is replaced by a continuous stable region before reaching the period-doubling bifurcation.

\section{\label{SIntro}Reinforcement Learning of Bipedalism}
\begin{figure}
	\centering
	\includegraphics[scale=1.2]{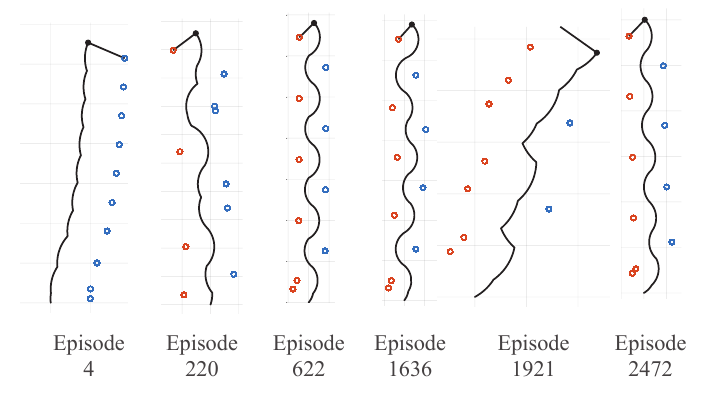}
	\caption{\textbf{Learning progress}. Training progress of one instance of the reinforcement learning agent for the little skate model with center of mass trajectories and footsteps at different episodes during learning.  }
	\label{Fig5}
\end{figure}

Our analysis so far shows that aquatic bipedalism has few requirements towards morphology (rudimentary legs) and control (constant torque, touchdown angle and frequency), but we have imposed an alternating leg sequence with fixed foot placement angle and torque for the analysis. We have not addressed the question if this gait can be discovered and if it is optimal when these constraints are relaxed. Discovery in this context requires an aquatic organism with rudimentary appendages to learn the neural control for a bipedal gait, which may be hard if the gait is a ``needle in a haystack'' in the vast control space. The search for a favorable gait relates to the field of gait selection and optimization \cite{hoyt1981gait,ruina2005collisional,srinivasan2006computer,alexander1980optimum}.   A natural question in the context of motor control learning is if a desired behavior can be acquired by reinforcement \cite{glimcher2011understanding}. With the objective of maximizing the travelled distance and minimizing the required energy (or equivalently, minimizing energy consumption for a travelled distance), we trained a reinforcement learning (RL) agent \cite{lillicrap2015continuous,sutton2018reinforcement} on the model to obtain a given locomotion speed $v_T$.  The framework  has four state parameters (planar position and velocity of the point mass) and four control parameters (three continuous ones for $T, \omega, \alpha_0$ and a binary one for the leg case $l$) - for details see Methods. We observed, in most instances of the learning routine like the one shown in Fig. \ref{Fig5}, that a one-legged locomotion strategy emerges after only 4 episodes, two-legged locomotion emerges around episode 200, and periodic locomotion with alternating leg cycles emerges via RL around episode 600. This type of gait prevails as the most efficient one and other explored strategies have a worse cumulative reward. We ran $\sim$ 50 instances of learning for 5000 episodes with changing learning parameters and weights for the reward function and found that the left-right  alternating gait emerged in 70\% of instances and generated the highest reward. The best solution matches the little skate's observed walking gait in Fig. \ref{Fig1}, and we see an undulatory motion of the center of mass and a left-right alternating leg sequence (see Fig. \ref{Fig2}C and SI video).
For comparison with the bifurcation diagram, the evolution of the best learned RL policy in the parameter space corresponds to the dashed lines in Fig. \ref{Fig3}, starting with no forward speed and $\gamma=1$. The first step uses a large $\gamma$ and over subsequent steps minimizes it while increasing $\bar{v}^*$, eventually settling close to the upper saddle-node bifurcation point at $\gamma=0$, confirming the optimality of the solution discovered by RL. In the context of our model, these results suggest that, despite the vast solution space of gaits, a left-right alternating bipedal control strategy can and will be discovered and is the optimal solution for energy efficient locomotion.

\section{\label{SIntro}Robotic Model of Bipedalism}
\begin{figure*}
	\centering
	\includegraphics[scale=0.55]{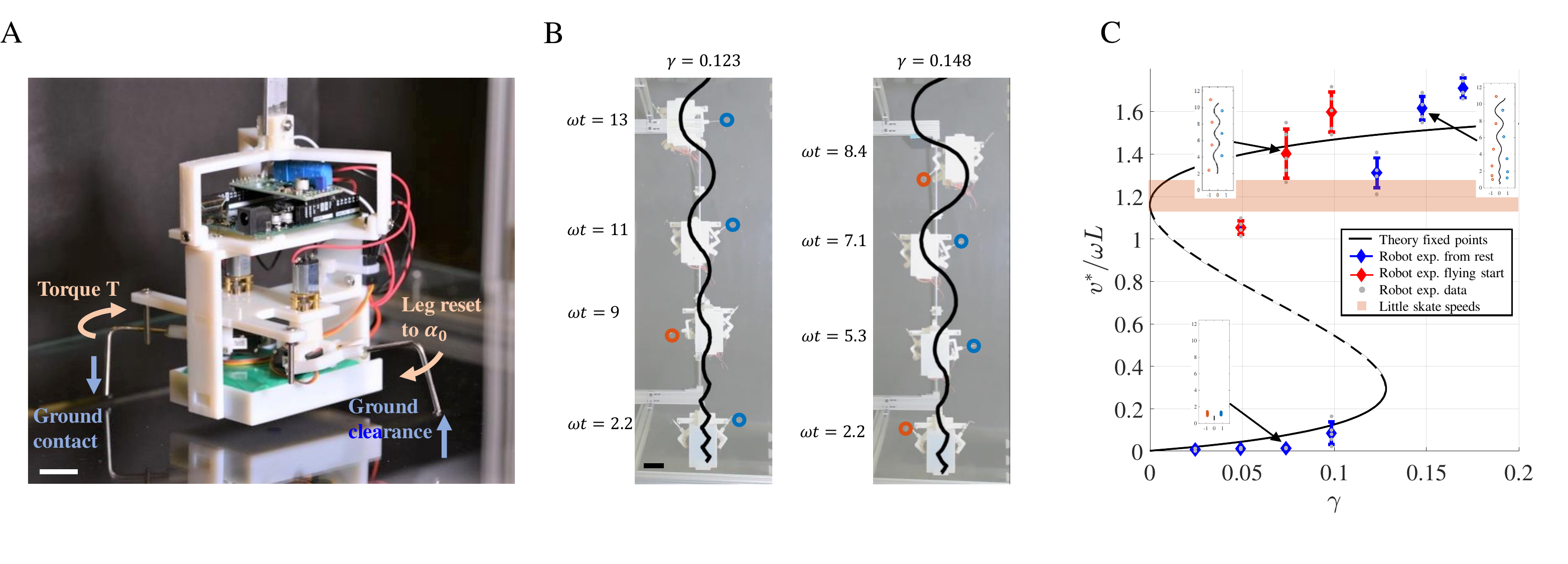}
	\caption{\textbf{Robot experiments.} \textbf{A}, Image of supported robot with right leg in contact with the ground and left leg resetting to initial angle $\alpha_0$. Scale bar 5cm. \textbf{B}, Bottom-up view of sequence of robot walking gait for two  cases initialized from rest. The black line indicates the center of mass trajectory of the robot. Circles indicate closed leg contact points for corresponding picture. Scale bar 5cm. \textbf{C}, Experimental fixed-point velocities as a function of nondimensional torque $\gamma$. The shaded region is the range of observed little skate speeds as obtained from the video analysis in \cite{jung2018ancient}. Insets show a selection of experimental trajectories of the center of mass (black line) as a function of dimensionless position with footsteps (circles).}
	\label{Fig6}
\end{figure*}

Having shown that the bipedal control strategy is the preferred solution in an optimization framework, we turn to realize it experimentally, inspired by a range of recent studies in this vein such as mimicking the legged locomotion of mudskippers \cite{mcinroe2016tail}, the reconstruction of feasible tetrapod gaits in extinct species \cite{nyakatura2019reverse}, the cost of transport of high frequency swimming \cite{zhu2019tuna}, and robophysical studies in general \cite{aguilar2016review}.   We used a supported (simulating neutrally buoyant environments) robotic biped as shown in Fig. \ref{Fig6}A. The legs were mechanically constrained to satisfy $\alpha \in [0, 2.15] rad$ and we fixed $\omega$,$L$,$m$ and varied $T$ to change $\gamma$. The design of the robot aims to test the planar dynamics of aquatic walking, restricting vertical oscillations of the body (but not the vertical displacement of legs) for simplicity; an unsupported system would require stabilization of vertical body attitude and position by e.g. using a tail or pectoral fins that generate lift. We also ignored the effect of fluid drag for the study of planar dynamics as we found no qualitative change in the bifurcation diagram when fluid drag was considered.

Fig. \ref{Fig6}B shows a bottom view of a series of snapshots at different times of two experiments. The system was initialized from rest and with $\gamma=0.123$ and $\gamma=0.148$. The solid line corresponds to the center of mass trajectory of the robot and the dots to the footsteps of the snapshots. The coexistence of fixed points at $\gamma<0.13$ was tested by initializing the robot from rest and with a flying start, i.e. an initial nonzero velocity. As expected we observed two steady solutions: a slow and a fast gait. Note, however, as $\gamma$ increases, we no longer require a nonzero velocity initialization to reach the fast solution branch, effectively demonstrating that gait transition (resting to walking), acceleration, and stabilization are performed without the need for additional control. As observed in skate experiments and our model, the robot also exhibits undulatory behavior and a regular footstep pattern. The observed versus predicted locomotion speeds are shown in Fig. \ref{Fig6}C. The observed locomotion speeds are low when started from rest for $\gamma<0.13$, but converge to the upper branch of the bifurcation diagram for $\gamma>0.13$, with exception of the $\gamma=0.123$ case where a convergence to the upper branch occurs at marginally lower $\gamma$ than predicted. All together, the gait is completely determined in terms of a constant rate of motion $\omega$, range of motion $\alpha_0$, and energetic cost as determined by the constant torque $T$.

\section{\label{SIntro}Discussion}
Under the assumptions of neutrally buoyant environments and rudimentary leg-like appendages we showed that a left-right alternating bipedal locomotion strategy exists in a minimal model and that it is stable, energy efficient, learnable, and easily realizable in a robotic platform. The fact that such a simple control strategy leads to remarkably robust and efficient behavior is related to the concept of passive dynamics, as seen for instance in a slinky toy ``walking'' down a slope without the need of complex control and the passive dynamic walkers \cite{mcgeer1990passive} that revolutionized the way we think about human-like locomotion. Such systems demonstrate that the appropriate morphology for a particular environment often leads to the most efficient behavior with remarkable simple or no active feedback control. In the same vein, it was shown that anesthetized fish can swim upstream given the right environmental conditions \cite{beal2006passive} which revealed that the concept of passive dynamics is indeed exploited in natural systems. The little skate robot presented here is not passive as the energy source for motion is an internal actuator and not an external source of energy like gravitational potential energy, but it exploits the same principles of a passive system: sustained locomotion under a constant energy source without feedback control. 

The selection of the walking gait in the little skate may be a consequence of an increasing energetic cost of transport for swimming at slow locomotion speeds \cite{di2016skating}, similar to the gait transition in other vertebrates \cite{hoyt1981gait}. Metabolic rate measurements of walking skates are yet to be recorded and will provide further insights into energy expenditure as a driver of gait selection, but the passive bipedal gait presented here can help explain the energetic benefits of benthic walking in aquatic environments. It must be added that the little skate uses an alternative legged gait called punting \cite{koester2003punting} which it uses, for example, to kick-start the left-right alternating locomotion strategy. Punting uses two legs simultaneously and was not discovered in our optimization framework, but it may be the preferred gait when fast acceleration is rewarded over energy efficient locomotion.

% zero energy walking and other models
Our study complements earlier work on the theoretical existence of zero-energy gaits in terrestrial walking\cite{gomes2011walking} by showing how it arises in a minimal theoretical model for aquatic bipedalism, and approximately in robot experiments. In particular, it requires the legs to be collinear at the end/beginning of every footstep, effectively reducing the degrees of freedom of the problem. Instead of controlling the legs individually, one leg can simply mirror the motion of the other leg, reminiscent of mirror algorithms used in other impulsive robotic tasks like juggling \cite{buhler1990family}.  This type of gait can also be realized using a rigid body with two attached rigid legs;  walking then corresponds to alternate rotations about a vertical axis, centered about one of two legs. This is similar to the waddling gait of penguins, where lateral undulation is thought to improve the energetics of locomotion \cite{griffin2000biomechanics}. Of course such zero-energy models do not account for losses due to internal damping, cost of leg swing, or contribution of leg mass to collision,  fluid drag etc. Adding fluid drag to our model, we found no qualitative difference in dynamics. Comparing the observed steady locomotion speed in the little skate  $v^*\in[1.1,1.26]$, we found that it is generally lower than our measured speeds in robots, and might correspond to the gait with no energy loss (see Figs. \ref{Fig1}D, SI).

Together, our results demonstrate the minimal requirements in a neural control strategy (constant force input, stability, learnability)  while obtaining high energy efficiency (zero-cost gait). The minimal model of aquatic bipedalism that we present yields  a stable, open-loop, energy-efficient gait that can be learned using a minimal reinforcement scheme, and can be physically realized in robot experiments. It is also consistent with biological observations in the little skate. Our study also reinforces how the physical environment, the morphology of the organism and the neural substrates synergistically produce a coordinated walking gait, linking to fundamental questions in passive dynamics, self-organization, and embodiment \cite{pfeifer2007self}.  While we may never know exactly how the first walking gait arose, the combination of the embodied passive dynamics associated with a minimal legged morphology that are ancient \cite{standen2014developmental}, and the presence of conserved neural circuits that are now known to be equally ancient \cite{jung2018ancient}  may well have helped pave the way for legged-gaits before our aquatic ancestors transitioned to {\it terra firma}. Understanding how the brain, body and environment worked together in heterogeneous aquatic and terrestrial environments likely needed to include proprioceptive feedback. But in reliably homogeneous environments, perhaps the simple strategy quantified here was where it all started.

%In this spirit, the robotic skate presented here illuminates physical limits of our model and provides testable values for torques and cost of transport in biomimetic aquatic walking.

\section{\label{SIntro}Methods}

\textbf{Animal data}. Kinematic data from little skates was obtained from supplementary movies in \cite{jung2018ancient} with permission from the authors. Center of mass position, body orientation, and leg positions were extracted using the software Kinovea. Some characteristics of the extracted data are shown in Table \ref{ETab1}. The animals were tested in a tank with a textured PDMS surface for traction of the legs with the substrate. Slip per step of the leg during stance phase varied across individuals and ranged between $0.1$mm to $1$mm which corresponds to 0.5\% to 5\% of the step length. Angle plots of $\alpha$ (Figs. 1D, SI) were obtained from measuring the angle between the centerline (from tracking the connecting line between pelvic girdle and mouth) and vectors pointing from the pelvic girdle to the leg tips. Velocities of pelvic girdle as a function of time (insets in Figs. 1D, SI) were computed from filtered trajectories (local regression using weighted linear least squares and a 2nd degree polynomial model using a span of 10\% of the total number of data points) and numerically differentiating them with respect to time. Data was made dimensionless with a leg length $L=1.15cm$ and a frequency $\omega=1.1Hz$ which were extracted from the movies.  

\begin{table} [h]
\caption{\textbf{Mean (bold) and standard deviation (parentheses) of kinematic data from 3 individual skates}. Data averaged over 10 steps in experiment excluding initial acceleration. $v_m$ mean nondimensional velocity, $\phi$ phase difference of legs, $\alpha_p$ peak leg angle, slip/step normalized by leg length.}
\begin{center}
\begin{tabular}{ |c||c|c|c|c| }
\hline
Skate & $v_m$ & $\phi$ &$\alpha_{p}$ & slip/step \\
\hline\hline
1     &         $\bm{1.3} (0.2)$       &     $\bm{162^{\circ}}(34^{\circ}) $     &       $\bm{2.19}(0.7)$        &            $\bm{0.088}(0.1)$                 \\
\hline
2     &        $\bm{1.38}(0.16)$       &     $\bm{170^{\circ}}(10^{\circ}) $     &       $\bm{2.21}(0.1)$         &            $\bm{0.12}(0.14)$                 \\
\hline
3     &      $\bm{1.24}(0.3)$       &     $\bm{186^{\circ}}(10^{\circ}) $     &      $\bm{2.33}(0.02)$            &            $\bm{0.006}(0.06)$                 \\
\hline            
\end{tabular}
\end{center}
\label{ETab1}
\end{table}

\textbf{Reinforcement learning framework.}
For the model-based optimization of the little skate gait we used a reinforcement learning (RL) framework due to the obvious links between episodic and biological learning. Other optimization methods such as trajectory optimization can also find the optimal solution, but would not provide insight into the learnability of the walking gait via processes related to biological reinforcement \cite{glimcher2011understanding}. We chose a deep deterministic policy gradient (DDPG) reinforcement learning agent for the optimization of the little skate gait. DDPG \cite{lillicrap2015continuous} has the advantage of accepting continuous control inputs, which is commensurate with the biological control capabilities of the little skate. The dynamics for the RL environment are obtained by computing the next step position after placing leg $l$ at angle $\alpha_0$ on the ground and applying a torque $T$ for $1/\omega$ seconds. This provides the new position coordinates $x,y$ and velocities $\bm{v}=(\dot{x},\dot{y})^T$. We ignored the rotational degree of freedom of the little skate center of mass for simplicity. At every episode, the center of mass is placed at the origin and its velocity set to zero. The reward of step $i$ was defined as
\begin{equation}\label{eq:reward}
    R_i = - |v_y-v_T | + \Delta y \left(1 - \frac{T}{T_{max}}. \right)
\end{equation}
The first term on the left side penalizes variations of the end-of-step vertical component of the velocity from the target speed $v_T$. This term drives the system towards a constant locomotion speed $v_T$. The second term accounts for optimization of the cost of transport, in that it rewards the product of travelled distance $\Delta y$ and negative normalized torque. $T_{max}$ is the maximum applicable torque in the system defined as a bound in the RL problem. The bounds for control parameters were $T\in[0, 1], \omega \in [1, 1000], \alpha_0 \in [0, \pi]$, and $l \in \{-1, 1\}$. We used Matlab's reinfrocement learning toolbox to train the critic and actor networks with two fully connected layers with 400 and 300 nodes and rectifiers as activation functions (except for the actor ouput where we used a tanh function). The leg case (-1 left, 1 right) was treated as a continuous control variable which was put through a signum function before its use. To test the effects of learning parameters on the converged solution we ran combinations of values for noise variance $\{0.1,0.2,0.3\}$, discount factor $\{0.8,0.9,0.99\}$ and learning rates $\{5\times 10^{-2},5\times 10^{-3},5\times 10^{-4}\}$. We ran the RL routine for 5000 episodes (an episode was ended after a maximum of 30 steps or if the center of mass surpassed the bounds at $x=\pm l$) for all combinations of parameter values and found convergence to the optimal bipedal gait in 17 of 27 cases, one of them is shown in Fig. \ref{Fig5} (all routines with learning rates of $5\times 10^{-2}$ did not converge). We further asked, whether changing the relative weight of the two terms in the reward function (\ref{eq:reward}) had an effect on the optimal solution of the gait. We ran 20 learning routines of 5000 episodes each and weighted the terms 1:3, 1:1, and 3:1. The solution yielding the highest reward was again of the type shown in Fig. \ref{Fig5} (left-right alternating strategy) and was found in 16 of 20 cases. 

\textbf{Robot experiment.} We developed a supported legged robotic system to systematically test the model predictions. The robot body was created using PolyJet technology using VeroWhitePlus material. The robot is powered by a 6V Nickel-metal hydride battery and digitally controlled with an Arduino Uno microcontroller. A motor driver (pololu max14870) operated two 6V DC motors (pololu 50:1 micro metal gear motor medium power) to allow for leg rotation. A servo motor per leg ensured ground contact and clearance of the leg tips (Power HD Sub-Micro Servo HD-1440A). Small rubber pads were glued to the leg tips to reduce slip. Two pins were mounted to the robot structure which prevent the legs from exceeding the angle  mechanically, and the main robot structure prevented the leg angle from becoming negative, i.e. we always have $\alpha \in [0,2.15]$. The mass of the robot was $m=350g$ and leg length $L=8cm$. The robot was connected with a long and stiff aluminum bar to a ball bearing which moves on a linear 1m steel rod. The steel bar was mounted at an angle of $0.5^{\circ}$ at which the ball bearing could slide along the steel rod. This resulted in a decrease in height of the bar position in direction of travel of the robot. Although this decrease in gravitational potential along the bar could potentially be used as a source for acceleration of the robot, friction inside the ball bearing resulted in a marginal velocity loss if the system is started with an initial speed $v_0$. Note that this is a conservative set-up as our model predicts no velocity loss over time in case of no leg collisions with the ground. The robot is hanging above a glass plate 90cm in length onto which the feet could push against when activated to close ground contact. A webcam recorded the locomotion behavior from the bottom of the glass plate at 30fps and center of mass trajectories obtained by tracking a blue marker on the bottom of the robot were subsequently extracted by analyzing the videos using Matlab. See the SI for an illustration of the set-up.

The control strategy for walking was implemented as follows. Both legs are initialized at $\alpha_0=2.15$ before every trial. Leg switching frequency $\omega$ was set to 1.3Hz. At switching time both DC motors reverse their rotation direction and servo motors change their state from lifted to contact or from contact to lifted. The parameter $\gamma$ was tuned by changing the leg torque exerted by the DC motors, which was controlled by adjusting the supplied voltage set by the motor controller. See supplementary videos for various trials with a selection of $\gamma$'s.

The data generated for Fig. \ref{Fig6}C was obtained from 5 independent robot experiments per error bar. All experiments were initialized from rest except the three error bars on the upper branch in the bistable region which were initialized with a nonzero velocity. The nondimensional initial speed of all flying starts is $v=2.4$ on average with standard deviation $0.4$ which was large enough to push the system to a state which is attracted by the upper branch but not too large such that the speed can converge within the limited number of steps. The experiment duration was 20 steps and the velocities correspond to the one after the 20th leg transition or the last leg transition in the field of the camera’s view. Trials at $\gamma \approx 0.05$ which were started with nonzero velocity often converged to the lower branch and slow velocities. The results shown for this case are the 5 successful cases where terminal velocity in the camera’s field of view did not vanish. However, we cannot guarantee that these cases have converged or if they would further decay in a larger experimental set-up, which may explain the prediction error. In the case of $\gamma \approx 0.123$ we observe slower speeds than expected, which can be explained with the fact that the gait has not completely converged to the steady state speed. These long transient phases were observed in simulations where $\gamma$ is close but past the end of the bistable region, which corresponds well to the position of $\gamma \approx 0.05$.

\begin{acknowledgments}
We acknowledge support from the Swiss National Science foundation (F.G.) and by a MacArthur Fellowship (L.M.).
\end{acknowledgments}

\bibliographystyle{ieeetr}
\bibliography{references}
\end{document}

\title{Supplementary Figures for\\ ``Models of benthic bipedalism''}

\author{{ Fabio Giardina$^1$,  and L. Mahadevan$^{1,2}$}\\
{\small \em $^1$School of Engineering and Applied Sciences, Harvard University, Cambridge, Massachusetts 02138, USA \\
$^2$ Department of Physics, Department of Organismic and Evolutionary Biology, Harvard University, Cambridge, Massachusetts 02138, USA}}

\maketitle

\renewcommand{\theequation}{S\arabic{equation}}
\renewcommand{\thefigure}{S\arabic{figure}}
\renewcommand{\thetable}{S\arabic{table}}

\vfill
\noindent
\begin{minipage}{1.0\textwidth}
  \strut\newline
  \centering
  \includegraphics[scale=0.7]{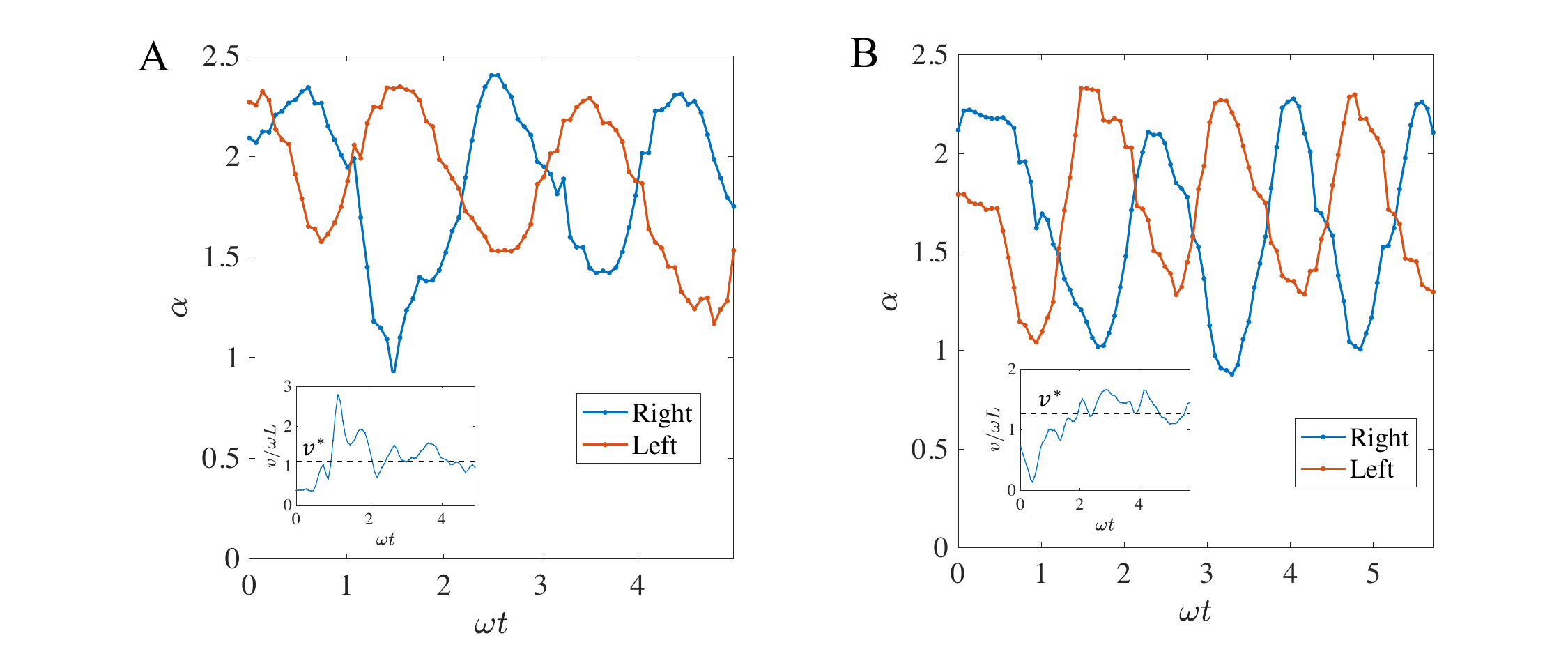}
  \captionof{figure}{\textbf{Locomotion kinematics of little skate A and B.}. These data are different specimen and runs from Fig. 1D. Left and right leg angles $\alpha$ as a function of dimensionless time and mean foot placement angle $\alpha_0$. The inset shows the dimensionless speed of the pelvic girdle as a function of dimensionless time with $v^*$  (dashed line) the approximate lower speed bound during steady state locomotion. Note that individuals A and B are different from the one corresponding to data in 1D.}
	\label{EFig1}
\end{minipage}

\vfill
\noindent
\begin{minipage}{1.0\textwidth}
  \strut\newline
  \centering
 \includegraphics[scale=0.6]{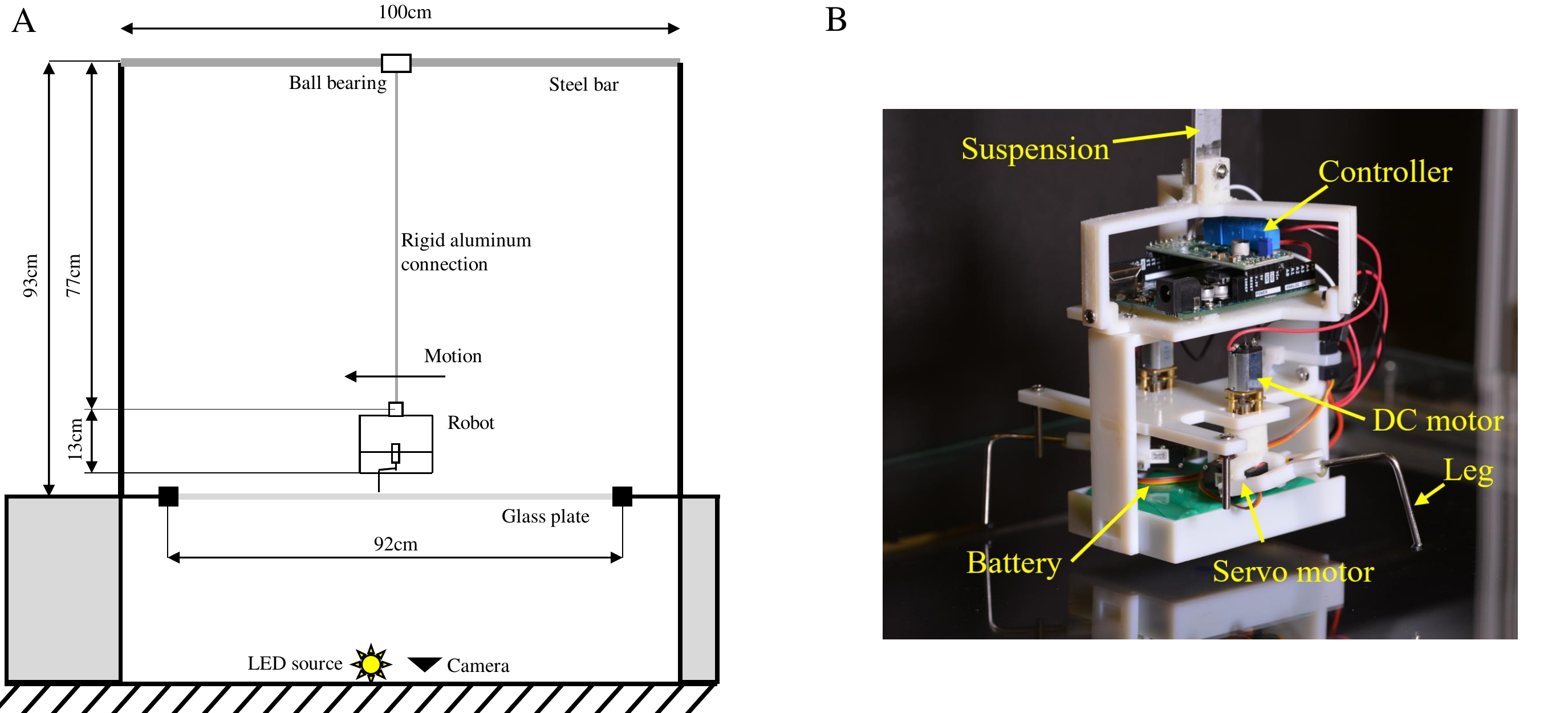}
  \captionof{figure}{\textbf{Sketch of experiment set-up (not to scale) and robot}. \textbf{A}. Robot is walking on a glass plate and filmed with a camera from below. A rigid aluminum beam supports the robot against gravity. The beam is connected via a ball bearing to a steel rod along which it can slide freely. \textbf{B} Robot walking on glass plate and its components.}
	\label{EFig3}
\end{minipage}

\clearpage

\begin{figure}
	\centering
	\includegraphics[scale=1.1]{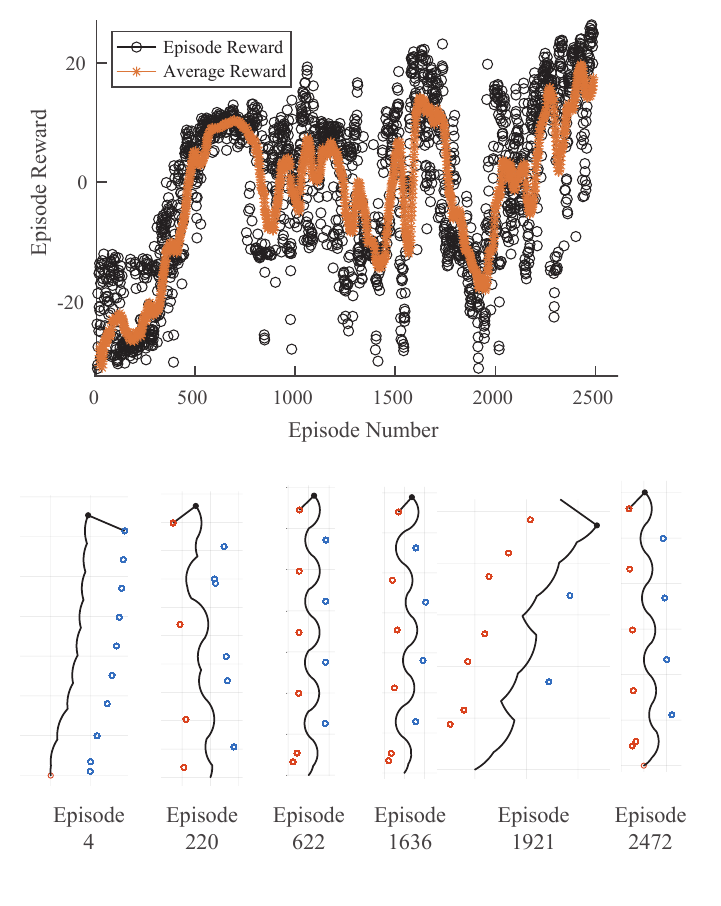}
	\caption{\textbf{Sample learning progress}. Training progress of DDPG agent for the little skate model with inset showing center of mass trajectories and footfalls at different episodes during learning.}
	\label{EFig5}
\end{figure}
\begin{figure}[H]
	\centering
	\includegraphics[scale=0.4]{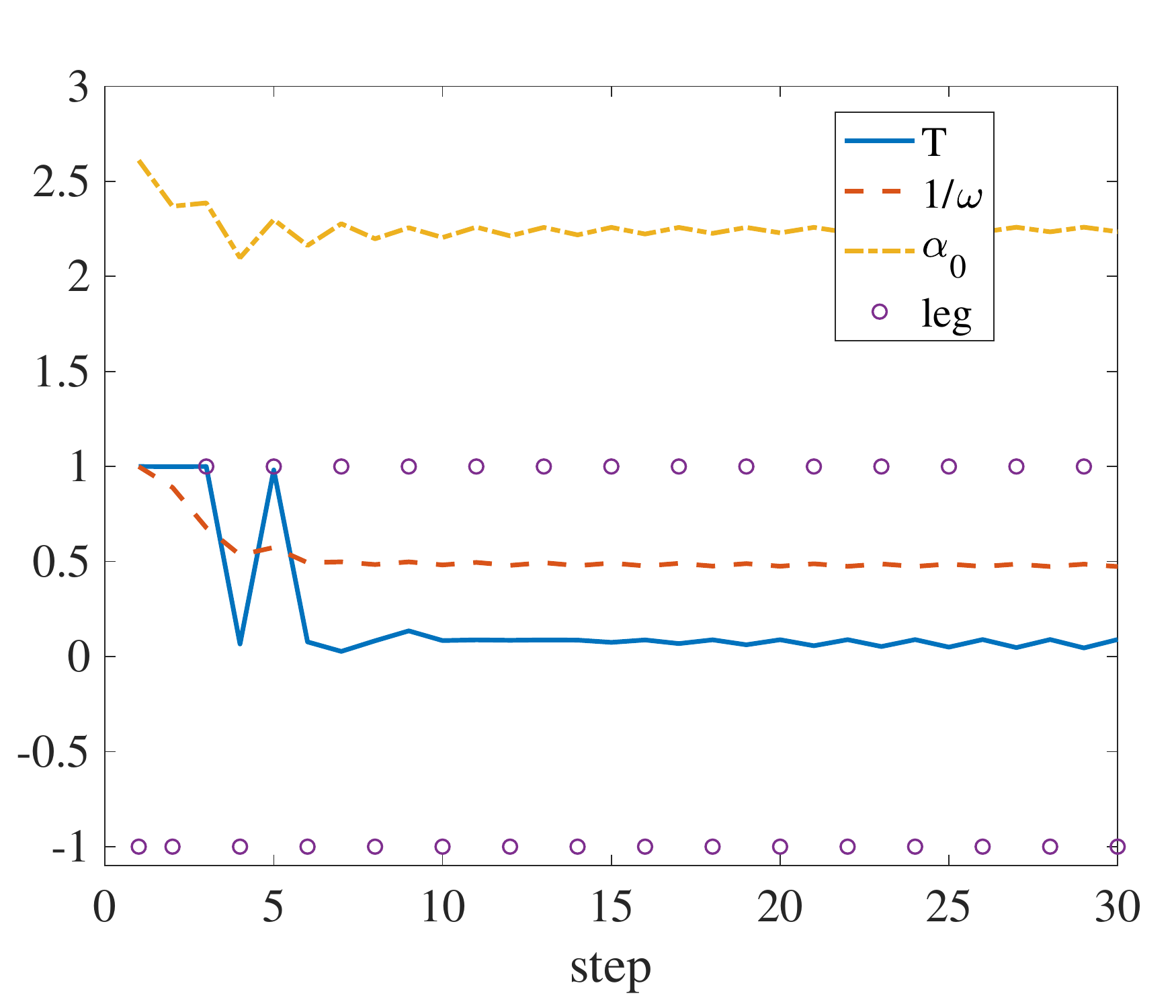}
	\caption{\textbf{Control parameters of learned solution.} Control parameters of best episode learned from the DDPG agent as a function of step number. This solution corresponds to RL path shown in Fig.3.}
	\label{EFig4}
\end{figure}

%\bibliographystyle{ieeetr}
%\bibliography{references}